\documentclass[reqno]{article}

\usepackage{graphicx}
\usepackage{amsmath,amsthm,amssymb}

\chardef\bslash=`\\ % p. 424, TeXbook
\theoremstyle{definition}

\theoremstyle{remark}

\newcommand{\eval}[2][\right]{\relax
  \ifx#1\right\relax \left.\fi#2#1\rvert}

\begin{document}
\title{\bf{An unusual eigenvalue problem}\footnote{It is a pleasure to dedicate
 this work to Professor Andrzej Staruszkiewicz on the occasion of his 65th birthday.}}

\author{Piotr Bizo\'n\\
Institute of Physics,
   Jagellonian University, Krak\'ow, Poland}
\maketitle
\begin{abstract}
\noindent We discuss an eigenvalue problem which arises in the
studies of asymptotic stability of a self-similar attractor in the
sigma model. This problem is rather unusual from the viewpoint of
the spectral theory of linear operators and requires special
methods to solve it. One of such methods based on continued
fractions is presented in detail and applied  to determine the
eigenvalues.

\end{abstract}
\section{Introduction}
Many nonlinear evolution equations have the property that
solutions which are initially smooth become singular after a
finite time. The nature of this phenomenon, usually called blowup,
has been a subject of intensive studies in many fields ranging
from fluid dynamics to general relativity. The problem whether the
blowup can occur, and if so, what is its character, is very
difficult for some major  evolution equations in physics, such as
Navier-Stokes  equations or Einstein's equations. Thus, in order
to get some insight, it seems useful to study toy models. This
paper is concerned with such a toy model, a nonlinear radial wave
equation
\begin{equation}\label{eq}
u_{tt} - u_{rr}-\frac{2}{r}u_r + \frac{\sin{(2u)}}{r^2}=0,
\end{equation}
where $r$ is the radial variable and $u=u(t,r)$. Equation
(\ref{eq}) describes equivariant wave maps from
  the $3+1$ dimensional Minkowski spacetime into
  the three-sphere (see \cite{bct} for the derivation).
In the
 physics literature this model is usually referred to as the sigma
 model.

The central question for equation (\ref{eq}) is whether
solutions starting from  smooth initial data
\begin{equation}
 u(0,r) = f(r), \qquad u_t(0,r) = g(r)
\end{equation}
may become singular in future? A hint towards answering this
question comes from the scaling argument. Note that equation
(\ref{eq}) is scale invariant: if $u(t,r)$ is a solution, so is
$u_{\lambda}(t,r)=u(t/\lambda,r/\lambda)$. Under this scaling the
conserved
 energy
\begin{equation}\label{energy}
E[u]= \int\limits_0^{\infty} \left(u_t^2+ u_r^2+
  \frac{ 2 \sin^2\!{u}}{r^2}\right) \: r^2 dr
\end{equation}
transforms as a homogenous function of degree one:
$E[u_{\lambda}]=\lambda E[u]$, which means that equation
(\ref{eq}) is supercritical in the language of the theory of
nonlinear partial differential equation. For supercritical
equations it is energetically favorable that solutions shrink to
small scales so singularities are expected to develop from
sufficiently large (in a suitable norm) initial data. Although
there are no rigorous results in this respect for equation
(\ref{eq}), an explicit example of a singularity forming from
smooth initial data is known. This example, first pointed out by
Shatah \cite{shatah} and later found in closed form by Turok and
Spergel \cite{ts} is provided by the self-similar solution
\begin{equation}\label{css0}
    u(t,r)=U_0(\rho)= 2 \arctan(\rho), \quad \text{where} \quad \rho=\frac{r}{T-t}
\end{equation}
and $T>0$ is a constant\footnote{$U_0$ is the ground state of a
countable family of self-similar solutions $U_n$ ($n=0,1,\dots$)
\cite{b}. However, all $n>0$ solutions are unstable so they do not
appear in the dynamics for generic initial data.}. Since
\begin{equation}\label{ssblow}
  \partial_r U_0(\rho)\Bigr\rvert_{r=0} = \frac{2}{T-t},
\end{equation}
the solution $U_0(\rho)$ becomes singular at the center when $t
\nearrow T$. By the finite speed of propagation, one can
    truncate this solution in space to get a smooth solution with compactly supported
    initial data which blows up in finite time.

In fact, the self-similar solution $U_0$ is not only an explicit
example of singularity formation, but numerical simulations
indicate that it appears as an attractor in the dynamics of
generic initial data \cite{bct}. We conjectured in \cite{bct} that
generically the asymptotic profile of blowup is universally given
by $U_0$, that is
\begin{equation}\label{conj}
 \lim_{t\nearrow T} u(t,(T-t) r) = U_0(r).
\end{equation}
To prove this conjecture one needs to understand the mechanism
responsible for the process of local convergence to the
self-similar solution $U_0$.
 Such a mechanism is relatively well-understood for nonlinear diffusion equations where the
global dissipation of energy is responsible for the convergence to
an attractor, however very little is known for conservative wave
equations where the local dissipation of energy is due to
dispersion.

In this paper, as the first step towards proving the above
conjecture, we describe in more detail how the limit (\ref{conj})
is attained. To this end, in section 2 we consider the problem of
linear stability of the solution $U_0$. This leads to an
eigenvalue problem which is rather unusual from the standpoint of
spectral theory of linear operators. In section 3 we solve this
problem  using the method of continued fractions. Finally, in
section 4 we present the numerical evidence
 that the deviation of the dynamical solution from the
self-similar attractor is asymptotically well described by the
least damped eigenmode.
\section{Formulation of the eigenvalue problem}
 In order to analyze the problem of linear stability of the self-similar solution $U_0$
 it is convenient to define the new time coordinate
$\tau=-\ln(T-t)$ and rewrite equation (\ref{eq}) in terms of
$U(\tau,\rho)=u(t,r)$
\begin{equation} \label{rho-tau}
U_{\tau\tau} + U_{\tau} + 2 \rho\: U_{\rho\tau}
-(1-\rho^2)(U_{\rho\rho} +\frac{2}{\rho} U_{\rho})
+\frac{\sin(2U)}{\rho^2}  = 0.
\end{equation}
In these variables the problem of finite time blowup in converted
into the problem of asymptotic convergence for $\tau \rightarrow
\infty$ towards the stationary solution $U_0(\rho)$.
 Following the standard procedure we seek
solutions of equation (\ref{rho-tau}) in the form
$U(\tau,\rho)=U_0(\rho)+ w(\tau,\rho)$. Neglecting the $O(w^2)$
terms we obtain a linear evolution equation for the perturbation
$w(\tau,\rho)$
\begin{equation}\label{pert}
w_{\tau\tau} + w_{\tau} + 2 \rho\: w_{\rho\tau}
-(1-\rho^2)(w_{\rho\rho} +\frac{2}{\rho} w_{\rho}) +\frac{2\cos(2
U_0)}{\rho^2} w  = 0.
\end{equation}
Substituting $w(\tau,\rho)=e^{\lambda \tau} v(\rho)/\rho$ into
(\ref{pert}) we get the eigenvalue equation
\begin{equation}\label{spectrum}
-(1-\rho^2) v''+2 \lambda \rho v' +\lambda(\lambda-1) v +
\frac{V(\rho)}{\rho^2} v=0,
\end{equation}
where
\begin{equation}\label{poten}
    V(\rho) = 2\cos(4\arctan\rho)= \frac{2
(1-6 \rho^2+\rho^4)}{\rho^2 (1+\rho^2)^2} .
\end{equation}
We consider equation (\ref{spectrum}) on the interval $0\leq
\rho\leq 1$, which corresponds to the interior of the past light
cone of the blowup point $(t=T,r=0)$. Since a
 solution of  the initial value problem for equation
(\ref{eq}) starting from smooth initial data remains smooth for
all times $t<T$, we demand the solution $v(\rho)$ to be analytic
at the both endpoints $\rho=0$ (the center) and $\rho=1$ (the past
light cone). Such a globally analytic solution of the singular
boundary value problem can exist only for discrete values of the
parameter $\lambda$, hereafter called eigenvalues.

A straightforward way to find the eigenvalues would be to use the
Frobenius method. The indicial exponents at the regular singular
point $\rho=0$ are $2$ and $-1$, hence the solution which is
analytic at $\rho=0$ has the power series representation
\begin{equation}\label{v0}
v_0(\rho) = \sum_{n=0}^{\infty} \alpha_n \rho^{2n+2}, \qquad
\alpha_0 \neq 0.
\end{equation}
Since there are no complex singularities in the open disk of
radius $1$ about $\rho=0$, the series (\ref{v0}) is absolutely
convergent for $0\leq \rho < 1$. At the second regular singular
point, $\rho=1$, the indicial exponents are $0$ and $1-\lambda$
so, as long as $\lambda$ is not an integer (below we shall discuss
this case separately), the two linearly independent solutions have
the power series representations
\begin{equation}\label{v12}
v_1(\rho) = \sum_{n=0}^{\infty} \beta_n^{(1)} (1-\rho)^n,\qquad
v_2(\rho) = \sum_{n=0}^{\infty} \beta_n^{(2)}
(1-\rho)^{n+1-\lambda}.
\end{equation}
These series are absolutely convergent for $0< \rho \leq 1$.
 If $\lambda$ is not an integer then
only the solution $v_1(\rho)$  is analytic at $\rho=1$. From the
theory of linear ordinary differential equations we know that the
three solutions $v_0(\rho)$, $v_1(\rho)$, and $v_2(\rho)$ are
connected  on the interval $0<\rho<1$ by the linear relation
\begin{equation}\label{connect}
    v_0(\rho) = A(\lambda) v_1(\rho) + B(\lambda) v_2(\rho).
\end{equation}
The requirement that the solution which is analytic at $\rho=0$ is
also analytic at $\rho=1$ serves as the quantization condition for
the eigenvalues $B(\lambda)=0$. Unfortunately, the explicit
expressions for the connection coefficients $A(\lambda)$ and
$B(\lambda)$ are not known for equations with more than three
regular singular points. In the next section we shall present an
indirect method which goes around this difficulty.

 We  remark in passing that alternatively the eigenvalues can be computed numerically
 using a shooting-to-a-midpoint technique. With this technique one approximates the solutions $v_0(\rho)$ and $v_1(\rho)$ by the
power series (\ref{v0}) and (\ref{v12}), truncated at some
sufficiently large $n$, and then computes the Wronskian of these
solutions at a midpoint, $\rho=1/2$, say. The zeros of the
Wronskian correspond to the eigenvalues. Although this technique
generates the eigenvalues with reasonable accuracy, it is
computationally very costly, especially for large negative values
of $\lambda$, because the power series (\ref{v0}) and (\ref{v12})
converge very slowly. Note also that shooting towards $\rho=1$
fails completely for large negative $\lambda$ because the solution
$v_2(\rho)$ is subdominant at $\rho=1$, that is, it is negligible
with respect to the analytic solution $v_1(\rho)$.
\section{Solution of the eigenvalue problem}
 In this section we shall solve
 the eigenvalue problem (\ref{spectrum}) using a method
 which exploits an intimate relationship between recurrence
 relations and continued fractions.
 Although this method is not widely
known, it is in fact quite old and has been applied in the past to
determine the bound states of the hydrogen molecule ion
\cite{jaffe} and quasinormal modes  of black holes \cite{leaver}.

  The key
idea is to determine the analyticity properties of the power
series solution $v_0(\rho)$ from the asymptotic behavior of the
expansion coefficients $\alpha_n$. In order to implement this idea
it is convenient to change the variables
\begin{equation}\label{subs}
    v(\rho)= (2-x)^{\frac{\lambda-1}{2}} y(x), \qquad x=\frac{2
    \rho^2}{1+\rho^2}.
\end{equation}
In terms of these variables equation (\ref{spectrum}) takes the
form
\begin{equation}\label{neweq}
    x^2 (1-x)(2-x) y'' + x [1-(1+\lambda) x(2-x)] y' -\frac{1}{4}
    [ \lambda^2 x(1-x)+9 x^2-17x+4] y = 0.
\end{equation}
The reason of making the transformation (\ref{subs}) is twofold.
 First, the transformation of the independent variable rearranges
 the singular points of equation (\ref{spectrum}) in such a way
 that, without changing the points $\rho=0$ and $\rho=1$,
moves the bothersome singularities
 at $\rho=\pm i$ (which lie on the unit disk around $\rho=0$ and obstruct the analysis of analyticity
  of the power series (\ref{v0}) at $\rho=1$) to infinity and moves $\rho=\infty$
 to $x=2$.
 Second, by factoring out the singular behavior at $x=2$ the number of terms in the recurrence relation
 for the coefficients of the  power series solution around $x=0$ is reduced from four to three.
The indicial exponents at $x=0$ are $1$ and $-1/2$ so the solution
which is analytic at $x=0$ has a power series expansion
\begin{equation}\label{y0}
y_0(x) = \sum_{n=1}^{\infty} a_n x^n, \quad a_1\neq 0.
\end{equation}
Substituting this series into equation (\ref{neweq}) we get the
three-term recurrence relation
\begin{eqnarray}\label{3term}
 &&p_2(0)  a_2 + p_1(0) a_1 =0, \nonumber \\
   &&  p_2(n) a_{n+2} + p_1(n) a_{n+1} + p_0(n) a_n
    = 0, \quad n=1,2,...
\end{eqnarray}
 with the initial conditions $a_0=0$ and $a_1=1$ (normalization)
where
\begin{eqnarray}\label{pn}
p_2(n) &=& 8 n^2 + 28 n + 20,\\
p_1(n) &=& -12 n^2 -(20+8\lambda) n -\lambda^2-8\lambda +9,\\
p_0(n) & =& 4 n^2 +4\lambda n +\lambda^2-9.
\end{eqnarray}
The series (\ref{y0}) is absolutely convergent for $0\leq x<1$ and
in general is divergent for $x>1$.
 In
order to determine the analyticity properties of the solution
$y_0(x)$ at $x=1$ we need to find the large $n$ behavior of the
expansion coefficients $a_n$. The three-term recurrence relation
(\ref{3term}) can be viewed as the second order difference
equation so it has two linearly independent asymptotic solutions
for $n\rightarrow\infty$.  Following standard methods (see, for
example, \cite{de}) we find
\begin{equation}\label{asym3}
a_n^{(1)} \sim n^{\lambda-2} \sum_{s=0}^{\infty}
\frac{c_s^{(1)}}{n^s} , \quad \text{and} \quad a_n^{(2)} \sim
2^{-n} n^{-\frac{3}{2}} \sum_{s=0}^{\infty} \frac{c_s^{(2)}}{n^s},
\end{equation}
where the coefficients $c_s^{(1,2)}$ can be determined recursively
from (\ref{3term}). Thus, in general the solution of the
recurrence relation (\ref{3term}) behaves asymptotically as
\begin{equation} \label{3termasym}
a_n \sim c_1(\lambda) a_n^{(1)}+c_2(\lambda) a_n^{(2)}.
 \end{equation}
  If the
coefficient $c_1(\lambda)$ is nonzero then
\begin{equation}
\frac{a_{n+1}}{a_n} \sim
\frac{a^{(1)}_{n+1}}{a^{(1)}_n}\rightarrow 1 \quad \text{as}\quad
n\rightarrow\infty,
\end{equation}
 hence the power series (\ref{y0}) is
divergent for $x>1$ (in fact, it has a branch point singularity at
$x=1$). On the other hand, if $c_1(\lambda)=0$ then
\begin{equation}
\frac{a_{n+1}}{a_n} \sim
\frac{a^{(2)}_{n+1}}{a^{(2)}_n}\rightarrow \frac{1}{2} \quad
\text{as}\quad n\rightarrow\infty,
\end{equation}
 and the
power series  (\ref{y0}) is absolutely convergent for $x<2$, in
particular the solution $y_0(x)$ is analytic at $x=1$.

 The
advantage of replacing the quantization condition $B(\lambda)=0$
in the connection formula (\ref{connect})  by the equivalent
condition
 $c_1(\lambda)=0$ follows from the fact that $c_1(\lambda)$ is the coefficient of
the dominant solution in (\ref{3termasym}), in contrast to
$B(\lambda)$ which is the coefficient of the subdominant solution
in (\ref{connect}). In the theory of difference equations
(recurrence relations) a subdominant solution, that is a solution
which is asymptotically negligible with respect to any other
solution, is called the minimal solution. In contrast to a
dominant solution, the minimal solution, if it exists, is unique.
The condition $c_1(\lambda)=0$ is thus equivalent to the
requirement that the solution of the recurrence relation
(\ref{3term}) starting with $a_0=0$ and $a_1=1$ is minimal. In
order to find when this minimal solution exists we shall use now
 a relationship
between three-term recurrence relations and continued fractions.
Let
\begin{equation}
    A_n=\frac{p_1(n)}{p_2(n)}, \quad B_n=\frac{p_0(n)}{p_2(n)},
    \quad r_n=\frac{a_{n+1}}{a_n}.
\end{equation}
 Then, we can rewrite (\ref{3term}) as
\begin{equation}
r_n = - \frac{B_n}{A_n+r_{n+1}},
\end{equation}
and applying this formula repeatedly we get the continued fraction
representation of $r_n$
\begin{equation}\label{pinch}
r_n = - \frac{B_n}{A_n -} \;\frac{B_{n+1}}{A_{n+1} -}\;
\frac{B_{n+2}}{A_{n+2} -} ...
\end{equation}
Pincherle's theorem \cite{de} says that the continued fraction on
the right hand side of equation (\ref{pinch}) converges if and
only if the recurrence relation (\ref{3term}) has a minimal
solution $a^{min}_n$ and, moreover, in the case of convergence,
equation (\ref{pinch}) holds with $r_n=a^{min}_{n+1}/a^{min}_n$
for each $n$.

 Using  Pincherle's theorem and setting $n=1$ in
 (\ref{pinch}) we obtain the eigenvalue equation
\begin{equation}\label{eigen}
\frac{a_2}{a_{1}} = \frac{1}{20} (\lambda^2+8\lambda-9)= -
\frac{B_{1}(\lambda)}{A_{1}(\lambda) -}\;
\frac{B_{2}(\lambda)}{A_{2}(\lambda) -}...
\end{equation}
The continued fraction in (\ref{eigen}), which by Pincherle's
theorem is convergent for any $\lambda$, can be approximated with
essentially arbitrary accuracy  by downward recursion starting
from a sufficiently large $n=N$ and some (arbitrary) initial value
$r_N$. The roots of the transcendental equation (\ref{eigen}) are
then found numerically (see table 1).
\begin{table}[h]
\centering
$$
\begin{tabular}{|c|cccccc|} \hline
$n$ & $0$ & $1$ & $2$ & $3$ & $4$ & $5$\\
& & & & & & \\
 $\lambda_n$ & 1 &  -0.542466  & -2  & -
3.398382 & -4.765079& -6.102295\\
 \hline \hline

$n$ & $6$ & $7$ & $8$ & $9$ & $10$ & $11$\\
& & & & & &  \\
$\lambda_n$ & -7.297807 & -7.765347 & -8.853889& -10.1228208 & -11.196495 & -11.802614\\
 \hline
\end{tabular}
$$
\caption{The first twelve  eigenvalues.}
\end{table}

 A glance  at table 1 shows some interesting properties of the
spectrum. First, all the eigenvalues are real. Second, there are
no eigenvalues  $\lambda>1$. Third, there are two integer
eigenvalues $\lambda=1$ and $\lambda=-2$.  Below we discuss these
properties in detail.
\subsubsection*{Why the eigenvalues are real?}
We find this property  surprising because we cannot see any a
priori reason which forbids complex eigenvalues. On the contrary,
it is easy to deform the potential $V$ in equation
(\ref{spectrum}), without changing the character of singularities
at the endpoints, in such a way that the resulting eigenvalues are
complex. Thus, the reality of eigenvalues is related to the
special form of the potential $V$. It remains a puzzle whether
this relationship is accidental or there is something deep to it.
\subsubsection*{The eigenvalue $\lambda=1$}
This eigenvalue is due to the freedom of changing the blowup time
$T$. To see this, consider a solution $U_0(\frac{r}{T'-t})$ with
shifted blowup time. In terms of the similarity variables
$\tau=-\ln(T-t)$ and $\rho=r/(T-t)$ we have
\begin{equation}\label{zeromode}
U_0\left(\frac{r}{T'-t}\right)= U_0\left(\frac{\rho}{1+\epsilon
e^{\tau}}\right), \quad \text{where} \quad \epsilon=T'-T.
\end{equation}
Thus, the perturbation induced by the shift of blowup time has the
form
\begin{equation}\label{shift}
    w(\rho,\tau) = -\epsilon e^{\tau} U'_0(\rho),
\end{equation}
and consequently the mode
\begin{equation}\label{vgauge}
    v(\rho) = \rho U'_0(\rho) = \frac{\rho^2}{1+\rho^2}
\end{equation}
solves equation (\ref{spectrum}) with $\lambda=1$. Since this mode
is evidently analytic at both $\rho=0$ and $\rho=1$, we conclude
that $\lambda=1$ is an eigenvalue. We emphasize that this positive
eigenvalue  should not be interpreted as the physical instability
of the solution $U_0$, as it is an artifact of introducing the
similarity variables and does not show up in the dynamics for
$u(t,r)$.
\subsubsection*{Nonexistence of eigenvalues $\lambda>1$}
 Using the transformation
\begin{equation}\label{trans}
v(\rho)=(1-\rho^2)^{-\frac{\lambda}{2}} z(\rho),
\end{equation}
we can put equation (\ref{spectrum}) into a standard
Sturm-Liouville form
\begin{equation}\label{selfad}
A z = \mu z, \quad \text{where} \quad A= -(1-\rho^2)^2
\frac{d^2}{d\rho^2} + \frac{1-\rho^2}{\rho^2} V(\rho), \quad
\mu=\lambda(2-\lambda).
\end{equation}
In the space of functions
$$
\mathcal{D} = L^2\left([0,1],\frac{d\rho}{(1-\rho^2)^2}\right)
$$
 the
operator $A$ is self-adjoint so the eigenvalues $\mu$ are real.
Both endpoints are of the limit-point type. Near $\rho=0$ an
admissible (that is, belonging to $\mathcal{D}$) solution behaves
as $z(\rho) \sim \rho$. Near $\rho=1$ two independent solutions
behave as
\begin{equation}\label{mu}
z_{\pm}(\rho) \sim (1-\rho)^{\frac{1}{2}(1\pm \sqrt{1-\mu})},
\end{equation}
 so only the solution $z_{+}$ with $\mu<1$ is admissible.
Now, we shall show  the operator $A$ has no eigenvalues. To see
this, note that the solution with $\mu=1$, corresponding to the
gauge mode
\begin{equation}\label{zeromode}
    z(\rho)= \sqrt{1-\rho^2} \frac{\rho^2}{1+\rho^2},
\end{equation}
has no zeros. This implies by a standard theorem for
Sturm-Liouville operators that there are no eigenvalues below
$\mu=1$. Since $\mu=1$ is not an eigenvalue (because the  mode
(\ref{zeromode}) does not belong to $\mathcal{D}$), we conclude
that the operator $A$ has the purely continuous spectrum $\mu\geq
1$.

What does this fact tell us about the eigenvalues of our problem?
Rewriting (\ref{mu}) in terms of $\lambda$ we have
\begin{equation}
z_{+}(\rho) \sim (1-\rho)^{\frac{\lambda}{2}}, \quad z_{-}(\rho)
\sim (1-\rho)^{1-\frac{\lambda}{2}},
\end{equation}
so comparing with (\ref{trans}) we see that only $z_{+}$ leads to
a solution $v(\rho)$ which is analytic at $\rho=1$. For
$Re(\lambda)>1$ the solution $z_{+}$ belongs to $\mathcal{D}$ so
in this case there is one-to-one relationship between the
eigenvalues of the operator $A$ and the eigenvalues of our
problem. Since the former has no eigenvalues, we infer that our
problem has no eigenvalues with $Re(\lambda)>1$. This result was
obtained previously in \cite{b}.

We point out that for $Re(\lambda)<1$ the requirements of
square-integrability  and analyticity near $\rho=1$ are mutually
exclusive so in this case there is no relationship between between
the eigenvalues of the operator $A$ and the eigenvalues of our
problem.
\subsubsection*{The algebraically special eigenvalue $\lambda=-2$}
If $1-\lambda=N$ is a positive integer then the two linearly
independent power series solutions of equation (\ref{neweq})
around $x=1$ are
\begin{equation}\label{y12}
y_1(x) = \sum_{n=0}^{\infty} b_n^{(1)} (1-x)^{n+N},\qquad y_2(x) =
C_N y_1(x) \ln(1-x) + \sum_{n=0}^{\infty} b_n^{(2)} (1-x)^n.
\end{equation}
The solution $y_1(x)$, corresponding to the larger indicial
exponent, is analytic at $x=1$ but the solution $y_2(x)$,
corresponding to the smaller indicial exponent, involves a
logarithm in general. However, for $N=3$ we have an exceptional
case:  the coefficient $C_3$  vanishes and  both solutions are
analytic at $x=1$. The best way to see this is to solve the
recurrence relation for the coefficients $b_n^{(2)}$ assuming
temporarily that the solution $y_2$ contains no logarithm. We find
\begin{equation}\label{ince}
    b_n^{(2)}=\frac{P_{2n-1}(\lambda)}{\lambda(\lambda+1)(\lambda+3)...(\lambda+n-1)},
\end{equation}
where $P_{2n-1}(\lambda)$ is a polynomial in $\lambda$ of order
$2n-1$. Moreover this polynomial has no integer roots. When
$1-\lambda=N$ is a positive integer different from $3$, then the
expansion  coefficient $b_N^{(2)}$ does not exist which
contradicts the assumption that $y_2$ contains no logarithm. The
only exception is $N=3$ because the denominator in (\ref{ince})
has no $(\lambda+2)$ term. Thus, for $\lambda=-2$ the singularity
at $x=1$ is  apparent - each solution which is analytic at $x=0$
is automatically analytic at $x=1$ as well. This proves that
$\lambda=-2$ is the eigenvalue.
\section{Numerical verification}
According to the linear stability analysis presented above  the
convergence of the solution $u(t,r)$ towards the self-similar
attractor $U_0$ should be described by the formula
\begin{equation}\label{numer}
u(t,r)=U(\tau,\rho) = U_0(\rho)+ \sum_{k=1}c_k e^{\lambda_k \tau}
v_k(\rho)/\rho \sim U_0(\rho)+c_1 e^{\lambda_1 \tau}
v_1(\rho)/\rho \quad \text{as} \quad \tau \rightarrow \infty,
\end{equation}
where $v_k(\rho)/\rho$ are the eigenmodes corresponding to the
eigenvalues $\lambda_k$ and $c_k$ are the expansion coefficients.
In order to verify (\ref{numer})  we solved equation (\ref{eq})
numerically for large initial data leading to blowup, expressed
the result in the similarity variables, and computed the deviation
from $U_0$ for $t \nearrow T$. The result (see figure 1) shows
that, in perfect agreement with the formula (\ref{numer}), the
deviation from $U_0$ is described by the least damped eigenmode
$v_1$. This makes us feel confident that the calculation presented
in section 3 contains no algebraic errors.
\begin{figure}[h!]
\centering
\includegraphics[width=\textwidth]{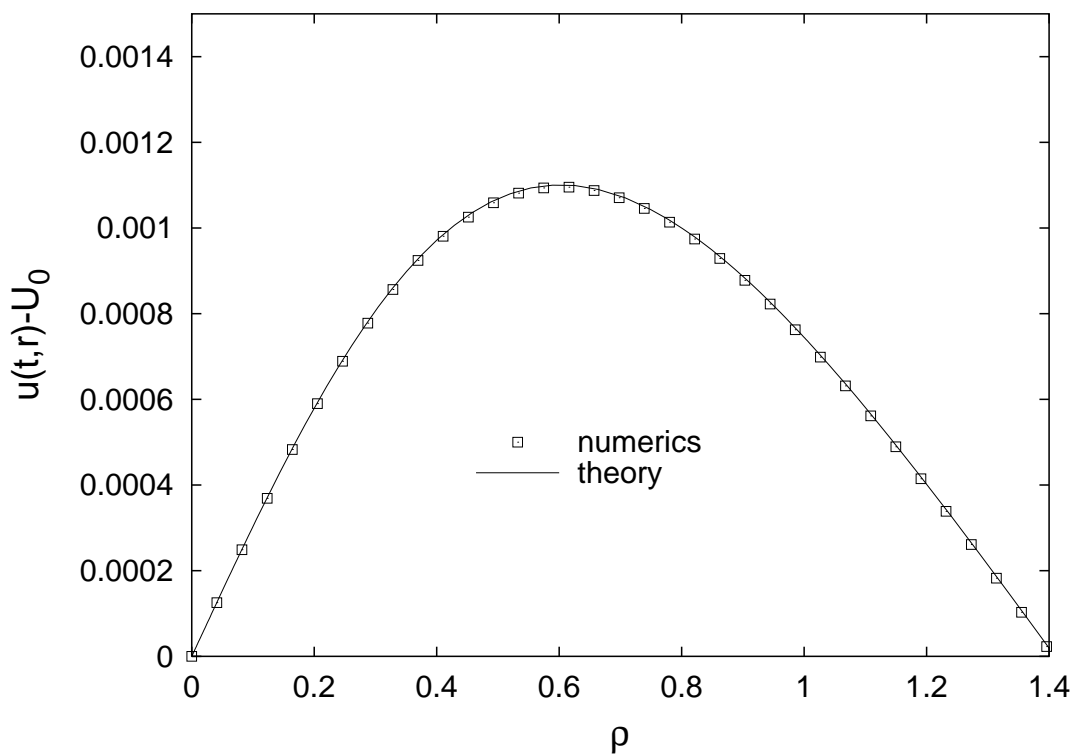}
\vskip 0.15cm
 \caption{\small{We plot the deviation of the dynamical solution
$u(t,r)$ from the self-similar solution $U_0=2 \arctan(\rho)$ at
some moment of time close to the blowup time. The solid line shows
the least damped eigenmode $c_1 (T-t)^{-\lambda_1}
v_1(\rho)/\rho$, where the coefficient $c_1$ is fitted once for
all times.}}
\end{figure}

\noindent We remark that the formula (\ref{numer}) could be used
to compute the eigenvalues numerically directly from the dynamics
rather than by solving the eigenvalue equation. Such a computation
 was performed by  Donninger \cite{roland} with the result which
is in rough agreement with table 1 (rough, as the dynamical
computation of eigenvalues is by far less accurate than the
continued fraction method).
\subsection*{Acknowledgments}
I thank Tadek Chmaj for remarks and providing  the numerical data
for section 4, and Roland Donninger for informing me about his
numerical results. This research was supported in part by the KBN
grant 2 P03B 006 23. The friendly hospitality of the Albert
Einstein Institute during part of this work is greatly
appreciated.


\begin{thebibliography}{10}


 \bibitem{bct} P. Bizo\'n, T. Chmaj, and Z. Tabor, Nonlinearity \textbf{13}, 1411 (2000).

 \bibitem{shatah} J. Shatah,
     Comm. Pure Appl. Math. \textbf{41}, 459 (1988).

\bibitem{ts} N. Turok and D. Spergel,  Phys. Rev. Lett. \textbf{64}, 2736 (1990).

\bibitem{b} P.~Bizo\'n,  Comm. Math. Physics \textbf{215}, 45 (2000).

 \bibitem{jaffe} G. Jaffe, Z. Phys. \textbf{87}, 535 (1934).

 \bibitem{leaver} E.W. Leaver, Proc. Roy. Soc. London
 \textbf{A402}, 285 (1985).

\bibitem{de} S. N. Elaydi, \emph{An Introduction to Difference
Equations} (Springer, 1999).

\bibitem{roland} R. Donninger, private communication.

\end{thebibliography}
\end{document}